\documentclass[preprint,12pt]{elsarticle}

\usepackage{graphicx}
\usepackage{amssymb}

\def\drbar{\ifmmode{\overline{\rm DR}} \else{$\overline{\rm DR}$} \fi}
\newcommand{\raw}{\rightarrow}

\title{SFOLD - a program package for calculating two-body sfermion decays at full one-loop level in the MSSM}

\address[HEPHY]{Institute of High Energy Physics, Austrian Academy of Sciences, A-1050 Vienna, Austria}%

\author[HEPHY]{H. Hluch\'{a}\corref{cor1}}
\ead{frisch@hephy.oeaw.ac.at}
\author[HEPHY]{H. Eberl}
\author[HEPHY]{W. Frisch}

\cortext[cor1]{Corresponding author}

\journal{Computer Physics Communications}

\begin{document}

\begin{frontmatter}

\begin{abstract}
  SFOLD (Sfermion Full One Loop Decays) is a Fortran program package for calculating all sfermion
  two-body decay widths and the corresponding branching ratios at full one-loop level within the MSSM. The package
  adopts the SUSY Parameter Analysis convention and supports the SUSY Les~Houches Accord input and
  output format. With the SFOLD package we found non-negligible electroweak corrections in bosonic decays
  of $\tilde b, \tilde t$ and $\tilde \tau$.
\end{abstract}

\begin{keyword}
Supersymmetry; Loop calculations; MSSM sfermion decays
\end{keyword}

\end{frontmatter}

\begin{small}
\noindent
{\em Manuscript Title:} SFOLD - a program package for calculating two-body sfermion decays
     at full one-loop level in the MSSM                                           \\
{\em Authors:} Hana Hluch\'a, Helmut Eberl, Wolfgang Frisch               \\
{\em Program Title:} SFOLD                                              \\
{\em Journal Reference:}                                                \\
{\em Catalogue identifier:}                                             \\
{\em Licensing provisions:} none                                        \\
{\em Programming language:} Fortran 77                                  \\
{\em Computer:} Workstation, PC                                         \\
{\em Operating system:} Linux                                           \\
{\em RAM:} 524288000                                                    \\ 
{\em Number of processors used:}                                        \\
{\em Supplementary material:}                                           \\
{\em Keywords:} Supersymmetry; Loop calculations; MSSM sfermion decays  \\
{\em Classification:} 11.1                                              \\
{\em External routines/libraries:} SLHALib 2.2                          \\
{\em Subprograms used:} LoopTools 2.6                                   \\
{\em Nature of problem:}\\
If the MSSM is realized in nature, LHC will produce
supersymmetric particles copiously. The best environment for a precise
determination of the model parameters would be a high energy  $e^+ e^-$ linear collider.
Experimental accuracies are expected at the per-cent down to the per-mill level.
These must be matched from the theoretical side. Therefore loop
calculations are mandatory.\\
{\em Solution method:}\\
This program package calculates all sfermion two-body decay widths and the corresponding
branching ratios at full one-loop level within the MSSM. The renormalization is done in the \drbar scheme
following the SUSY Parameter Analysis convention. The program supports the SUSY Les~Houches Accord
input and output format.
\\
{\em Restrictions:}\\
   \\
{\em Unusual features:}\\
   \\
{\em Additional comments:}\\
   \\
{\em Running time:} \\
\end{small}

\section{Introduction}

Supersymmetry (SUSY) is widely regarded as the most appealing extension of the
Standard Model (SM). Among supersymmetric theories, the Minimal Supersymmetric Standard Model
(MSSM) is the most extensively studied. If the MSSM is realized in nature, LHC will produce
supersymmetric particles copiously. The best environment for a precise
determination of the model parameters would be a high energy  $e^+ e^-$ linear collider.
Experimental accuracies are expected at the per-cent down to the per-mill level
\cite{R2,R3A,R2B}. These must be matched from the theoretical side. Therefore loop
calculations are mandatory.

There are a few program packages available for the automatic computation of amplitudes at full one-loop
level in the MSSM: FeynArts/FormCalc \cite{feynarts-formcalc}, SloopS \cite{SLOOPS_Higgs,SLOOPS_Sf}
and GRACE/SUSY-loop \cite{GRACE/SUSY-loop}. SloopS and GRACE/SUSY-loop also perform renormalization
at one-loop level. However, so far there is no publicly available code for the two-body sfermion decays
at full one-loop level within the MSSM. Therefore, we have developed the Fortran code SFOLD \cite{SFOLD}
(and HFOLD \cite{HFOLD-SFOLD, HFOLD}). Like HFOLD, it adopts the renormalization prescription of the
SUSY Parameter Analysis project (SPA) \cite{SPA} and supports the SUSY Les Houches Accord (SLHA) input and output
format \cite{SLHA1}. The package SFOLD (Sfermion Full One-Loop Decays) computes all two-body decay widths and
the corresponding branching ratios of all sfermions at full one-loop level.

Full one-loop radiative corrections to decays of sfermions into charginos and neutralinos are discussed
in \cite{ghs-fermionic-decays} for all sfermion flavours and generations.
Yukawa corrections to sbottom decay into lighter stop and charged Higgs boson are given in \cite{jinli}.
SUSY-QCD corrections to top and bottom squark decays into all Higgs bosons are calculated in \cite{sqcd1}.
SUSY-QCD corrections to stop and sbottom decays into weak bosons can be found in \cite{sqcd2}.
Finally, SUSY-QCD corrections to squark decays to gluinos are given in \cite{beenakker}.
Up to now, the electroweak corrections to sfermion decays into Higgs and gauge bosons have not been fully
addressed. It turns out that also these corrections cannot be neglected in a significant part of the parameter space.

\section{MSSM sfermion sector at tree-level}

\subsection{Masses and mixing angles}

The mass matrix in the gauge eigenstates basis $(\tilde f_L, \tilde f_R)$ is of the form
\cite{feynarts-MSSM}
\begin{equation}
  \mathcal{M}^2_{\tilde f} = \left( \begin{array}{cc}
  m^2_{\tilde f_L} & a_f^* m_f \\ a_f m_f & m^2_{\tilde f_R}
  \end{array} \right)
\end{equation}
where
\begin{eqnarray}
  m^2_{\tilde f_L} & = & M^2_{\tilde Q,\tilde L} + (I^{3L}_{f} - e_f s^2_W) \cos 2\beta M^2_Z + m^2_f \\
  m^2_{\tilde f_R} & = & M^2_{\tilde U,\tilde D,\tilde E} + e_f s^2_W \cos 2\beta m^2_Z + m^2_f \\
  a_f & = & A_f - \mu^* (\tan \beta)^{-2I^{3L}_f}
\end{eqnarray}
The mass eigenstates are obtained by diagonalizing $\mathcal{M}^2_{\tilde f}$ with a unitary matrix
$R^{\tilde f} = (\cos \theta_{\tilde f}, \sin \theta_{\tilde f}; -\sin \theta_{\tilde f},
\cos \theta_{\tilde f})$
\begin{equation}
  \textrm{diag} (m^2_{\tilde f_1},m^2_{\tilde f_2}) = R^{\tilde f}\mathcal{M}^2_{\tilde f}
  (R^{\tilde f})^\dag
\end{equation}
leading to the following sfermion masses and the mixing angle
\begin{eqnarray}
  m^2_{\tilde f_{1,2}} & = & \frac{1}{2} \left( m^2_{\tilde f_L} + m^2_{\tilde f_R}
  \mp \sqrt{(m^2_{\tilde f_L} - m^2_{\tilde f_R})^2 + 4 |a_f|^2 m_f^2} \right) \\
  \cos \theta_{\tilde f} & = & \frac{-a_f m_f}{\sqrt{(m^2_{\tilde f_L} - m^2_{\tilde f_1})^2}
  + |a_f|^2 m_f^2}
\end{eqnarray}
In the case of stop, sbottom and stau the left and right states are generally mixed. In contrast, sfermions from
first and second generation have negligible Yukawa couplings. Therefore, $\tilde f_1 = \tilde f_L$, $\tilde f_2 =
\tilde f_R$ if $(\mathcal{M}^2_{\tilde f})_{11} < (\mathcal{M}^2_{\tilde f})_{22}$ and $\tilde f_1 = \tilde f_R$,
$\tilde f_2 = \tilde f_L$ if $(\mathcal{M}^2_{\tilde f})_{11} > (\mathcal{M}^2_{\tilde f})_{22}$.

\subsection{Decay patterns}

There are four possibilities of Feynman graphs for a two-body decay of a scalar: the decay into two scalars,
into two fermions, into scalar and a vector particle and into two vector particles. The fourth possibility is
not realized in the decay of a sfermion in the MSSM. The following sfermion decays are calculated
(the first generation is shown, $i,j, c = 1,2;\, n = 1,\ldots, 4$):
\newline
\begin{center}
\begin{tabular}{|rcl|rcl|rcl|rcl|}
\hline
$\tilde \nu_e$ &$\raw$& $\nu_e \, \tilde \chi^0_n$ &
$\tilde e_i$   &$\raw$& $e     \, \tilde \chi^0_n$ &
$\tilde u_i$   &$\raw$& $u     \, \tilde \chi^0_n$ &
$\tilde d_i$   &$\raw$& $d     \, \tilde \chi^0_n$
\\
$\tilde \nu_e$ &$\raw$& $e     \, \tilde \chi^+_c$ &
$\tilde e_i$   &$\raw$& $\nu_e \, \tilde \chi^-_c$ &
$\tilde u_i$   &$\raw$& $d     \, \tilde \chi^+_c$ &
$\tilde d_i$   &$\raw$& $u     \, \tilde \chi^-_c$
\\
$\tilde \nu_e$ &$\raw$& $H^+ \, \tilde e_j$   &
$\tilde e_i$   &$\raw$& $H^- \, \tilde \nu_e$ &
$\tilde u_i$   &$\raw$& $H^+ \, \tilde d_j$   &
$\tilde d_i$   &$\raw$& $H^- \, \tilde u_j$
\\
$\tilde \nu_e$ &$\raw$& $W^+ \, \tilde e_j$ &
$\tilde e_i$   &$\raw$& $h^0 \, \tilde e_j$ &
$\tilde u_i$   &$\raw$& $h^0 \, \tilde u_j$ &
$\tilde d_i$   &$\raw$& $h^0 \, \tilde d_j$
\\
& & &
$\tilde e_i$   &$\raw$& $H^0 \, \tilde e_j$ &
$\tilde u_i$   &$\raw$& $H^0 \, \tilde u_j$ &
$\tilde d_i$   &$\raw$& $H^0 \, \tilde d_j$
\\
& & &
$\tilde e_i$   &$\raw$& $A^0 \, \tilde e_j$ &
$\tilde u_i$   &$\raw$& $A^0 \, \tilde u_j$ &
$\tilde d_i$   &$\raw$& $A^0 \, \tilde d_j$
\\
& & &
$\tilde e_i$   &$\raw$& $\tilde e_j \, Z$ &
$\tilde u_i$   &$\raw$& $\tilde u_j \, Z$ &
$\tilde d_i$   &$\raw$& $\tilde d_j \, Z$
\\
& & &
$\tilde e_i$   &$\raw$& $\tilde \nu_e \, W^-$ &
$\tilde u_i$   &$\raw$& $\tilde d_j   \, W^+$ &
$\tilde d_i$   &$\raw$& $\tilde u_j   \, W^-$
\\
& & &
& & &
$\tilde u_i$   &$\raw$& $u \, \tilde g$ &
$\tilde d_i$   &$\raw$& $d \, \tilde g$
\\
\hline
\end{tabular}
\end{center}
\vspace{0.5cm}
If the squark decay into a gluino is kinematically allowed it will dominate due to the QCD
interaction. The third generation $\tilde f_2$ can decay into $\tilde f_1$ and a neutral boson if there is
sufficiently large mass splitting. For stops and sbottoms with large mass differences, decays into
charged boson and a sfermion are possible.

\section{Calculation at full one-loop level}

The calculation of the decay widths is done in the same way as in the HFOLD program.
We work in the \drbar (dimensional reduction) renormalization scheme and in the general linear $R_\xi$
gauge for the $W^{\pm}$ and $Z^0$-boson. We wrote a Mathematica program that generated the whole Fortran
code using the packages FeynArts (FA) and FormCalc (FC). In FA all particle couplings in the MSSM are implemented.
We kept the divergent parts of the counterterms
to examine the UV finiteness of the renormalized amplitudes. The IR divergence is removed by using
soft bremsstrahlung or by adding a corresponding 1 to 3 process (hard bremsstrahlung) for which we
calculated all formulae analytically. The IR finiteness can be checked by varying the photon (gluon) mass
$\lambda$. Finally, we implemented several switchers that are described in sections 5.5, 5.6.

\section{Input parameters}

At the program start, SFOLD reads the file in SLHA format, where the Yukawa couplings, the gauge couplings
$g_1, g_2, g_3$, gaugino masses, the soft breaking terms, the VEV, $m_{A^{0}}, \tan \beta$, $\mu$ are taken
as input parameters at the scale Q. These parameters may be further changed. In that case, SFOLD recalculates
on-shell masses of Susy particles and does not take them from the input file.

\section{Program manual}

\subsection{Requirements}
\begin{itemize}
\setlength{\itemsep}{-1mm}
 \item {Fortran 77  (g77, ifort, gfortran)}
 \item{C compiler (e.g. gcc)}
 \item{LoopTools-2.6~\cite{looptools}}
 \item{SLHALib-2.2~\cite{slhalib}}
\end{itemize}

\subsection{About version 1.0}
\begin{itemize}
\setlength{\itemsep}{-1mm}
 \item{The CKM matrix is set diagonal}
 \item{Real SUSY input parameters}
 \item{One-loop corrections to Higgs masses}
 \item{Absence of three-particle sfermion decays}
\end{itemize}

\subsection{Installation}

\begin{enumerate}
\item Download the file \textbf{sfold.tar.gz} at
\begin{quote} http://www.hephy.at/tools \end{quote}
\item Unpack the archive by
\begin{quote} \tt{tar -xvzf sfold.tar.gz} \end{quote}
\item Go to the sfold folder and create symbolic links named LoopTools and SLHALib by
\begin{quote} \tt{ln -s ..path../LoopTools-m.n LoopTools} \\
              \tt{ln -s ..path../SLHALib-m.n SLHALib} \end{quote}
\item Then run
\begin{quote} \tt{./configure} \\
              \tt{make} \end{quote}
\item That will generate an executable called {\tt sfold}. To run SFOLD type
\begin{quote} \tt{./sfold} \end{quote}
\end{enumerate}

\subsection{Further notes}

\begin{itemize}
\setlength{\itemsep}{-1mm}
 \item{To use an older version of LoopTools you have to
 \begin{quote}
  - change the {\tt 'call ltini'} to {\tt 'call ffini'} in {\tt decay.F} file \\
  - include {\tt A00, A00C} in {\tt looptools.h} file
 \end{quote}}
 \item{LoopTools and SLHALib are installed correctly if folders {\tt ix86-linux/bin} and
       {\tt ix86-linux/include} were created. (The name {\tt ix86-linux} varies according
       to the system.) We further expect that the following files are present in the
       mentioned folders: {\tt looptools.h, libooptools.a, SLHA.h, libSLHA.a}.}
\end{itemize}

\subsection{The input file {\tt sfold.in}}

\begin{enumerate}
\setlength{\itemsep}{-1mm}
\item \textbf{name of input file (SLHA format)} \\
\item \textbf{type = 1,2,3,4,} \\ 1 = sneutrino, 2 = slepton, 3 = sup type, 4 = sdown type \\
\item \textbf{generation = 1,2,3} \\
\item \textbf{sfermion index = 1,2} \\
\item \textbf{bremsstrahlung = 0,1,2} \\0 = off, 1 = hard, 2 = soft  \\
\item \textbf{resummation of bottom yukawa coupling = 0,1} \\0 = off, 1 = on \\
\item \textbf{esoftmax} \\cut on the soft photon (gluon) energy if soft strahlung is used  \\
\item \textbf{name of output file}
\end{enumerate}

\subsection{The file {\tt sfold.F}}

Further parameters, switchers and options are:
%
%
\begin{center}
\begin{tabular}{ll}
\hline
{\tt delta\_in}          & divergent part of loop integrals \\
{\tt lambda\_in}         & photon (gluon) mass \\
{\tt Qscale}             & scale at which \drbar parameters are defined \\
{\tt xiW, xiZ}           & gauge parameters $\xi_W, \xi_Z$ \\
{\tt localchangesOn}     & If set to 0, on-shell masses are taken from SLHA \\
                         & input file. If set to 1, on-shell masses are calculated \\
                         & through self energies. (Must be set to 1, if some of \\
                         & the input parameters described in section 4 are \\
                         & changed.) Masses can be seen in {\tt masses.out} file. \\
{\tt osextmassesOn}      & If set to 0, \drbar masses in kinematics. If set to 1, \\
                         & on-shell masses in kinematics. \\
{\tt osloopmassesOn}     & If set to 0, \drbar masses in vertex corrections. \\
                         & If set to 1, on-shell susy masses in vertex corrections. \\
                         & (This is to avoid a trap when a process is not \\
                         & kinematically allowed with respect to \drbar masses.) \\
{\tt SMPRINTON}          & If defined, prints the SM parameters. \\
{\tt MSSMPRINTON}        & If defined, prints the MSSM parameters. \\
{\tt NO\_SQUARK\_MIXING} & If defined, squark mixing is switched off. \\
\hline \\
\end{tabular}
\end{center}
If any changes are done to {\tt sfold.F} file, it is necessary to run {\tt make} first.

\subsection{Working in Mathematica}

To work with SFOLD in Mathematica, the SPheno package (version 3.1.0) \cite{spheno} is required additionally.
To establish a link:
\begin{enumerate}
\item Compile the sfold with Mmakefile by
\begin{quote} {\tt make -f Mmakefile} \hspace{1cm} (or {\tt make -f Mmakefile-mac})\end{quote}
\item Open the {\tt sfold.nb} file
\item Go to the terminal and type
\begin{quote} {\tt ./Msfold -linkcreate \&} \end{quote}
\item Copy the output, return to {\tt sfold.nb} file and install link by
\begin{quote} {\tt link=Install[LinkConnect["...output..."]]} \end{quote}
\item Evaluate the cell. (It may be further necessary to go to the terminal and press Enter.)
      Continue working in the nb file.
\end{enumerate}
It is further neccesary to:
\begin{enumerate}
\item Specify the path to the SPheno executable
\begin{quote} {\tt SPhenoPath = "...path.../SPheno"} \end{quote}
\item Setting the Directory to sfold directory
\begin{quote} {\tt SetDirectory["...path.../sfold"]} \end{quote}
\end{enumerate}
We provided the user with six examples. The first one calculates the partial
widths of a sfermion at a specified mSugra point, the second one at a specified MSSM point,
the third one as a function of the mSugra parameter, the fourth one as a function of the
GMSB parameter, the fifth one as a function of the AMSB parameter and the sixth one as a function
of the MSSM parameter.
\newline
The input parameters are the same parameters as in {\tt sfold.in} and {\tt sfold.F} file except for the names
of the SLHA input and output file which are set to be {\tt SPheno.spc}, {\tt sfold.out}, respectively.
\newline \newline
The following functions and auxiliary lists are implemented:
\\
\begin{center}
\begin{tabular}{ll}
\hline
{\tt MakeSphenoSPCmsugra} & {\tt MakeSphenoSPCmsugra[s\_String, l\_List]} \\
{\tt msugraPara}          & \{m0, m12, tanb, sign(mu), A0\} \\
{\tt MakeSphenoSPCgmsb}   & {\tt MakeSphenoSPCgmsb[s\_String, l\_List]} \\
{\tt gmsbPara}            & \{Lambda, Mmes, tanb, sign(mu), n5, cgrav\} \\
{\tt MakeSphenoSPCamsb}   & {\tt MakeSphenoSPCamsb[s\_String, l\_List]} \\
{\tt amsbPara}            & \{m0, m32, tanb, sign(mu)\} \\
{\tt MakeSphenoSPCmssm}   & {\tt MakeSphenoSPCmssm[s\_String, l\_List]} \\
{\tt mssmPara}            & \{Q,M1,M2,M3,Au,Ad,Ae,Ac,As,Amu,At,Ab, \\
                          & Atau,Mue,MA02,TB,MSL1,MSL2,MSL3,MSE1, \\
                          & MSE2,MSE3,MSQ1,MSQ2,MSQ3,MSU1,MSU2, \\
                          & MSU3,MSD1,MSD2,MSD3\} \\
{\tt Decay}               & Decay[l\_List] \\
{\tt inputPara}           & \{type,gen,index,bremsOn,resumOn,esoftmax, \\
                          & Qscale,delta\_in,lambda\_in,xiZ,xiW,localchangesOn, \\
                          & osextmassesOn,osloopmassesOn\} \\
{\tt GetDM}               & {\tt GetDM[l\_List]} \\
\hline \\
\end{tabular}
\end{center}
{\tt MakeSphenoSPCmsugra[], MakeSphenoSPCgmsb[], MakeSphenoSPCamsb[]}
\newline create the file SPheno.spc. At first,
LesHouches.in file is created and then SPheno is called. {\tt s\_String} is the path to the SPheno executable,
{\tt l\_list} is the list of the corresponding model parameters (see {\tt msugraPara, gmsbPara, amsbPara} above).
\newline
{\tt Decay[]} calculates the partial widths of the specified sfermion. The rows correspond to decay modes, the first
column is the tree level result, the second is SUSY-QCD and the third one is the full result. For the input parameter
list, see {\tt inputPara} above.
\newline
{\tt GetDM[]} returns the particular decay mode of the specified particle. {\tt l\_List} consists of the
particle type, particle generation, particle index and of the decay mode index.

\section{Electroweak corrections in bosonic decays}

We give a few examples using the SFOLD package.
First, we focus on the $\tilde b_2$ decay. Figure \ref{sbottom2_t3A0} shows the partial widths as a function of
the mSugra parameter $m_0$. Other parameters are: $m_{\frac{1}{2}} = a m_0^2 + b m_0 + c$, $\tan\beta = 3$,
$\textrm{sign} (\mu) = 1$, $A_0 = 0$, where $a, b, c$ are chosen such that the whole parabola lies just above the
excluded region in Figure 2 of reference \cite{atlas}. The parabola goes through the points
[40, 330], [450, 300] and [740, 120]. For Figure \ref{sbottom2_t10A-300} we took $\tan\beta = 10$ and
$A_0 = -300$. Here the dominant decay is the decay into $\tilde t_1$ and $W^-$ if $m_0 \lesssim 200$.

{
\setlength{\unitlength}{1cm}
\begin{figure}[htbp]
\centerline{
\begin{picture}(13,7.3)
\includegraphics[width=12cm]{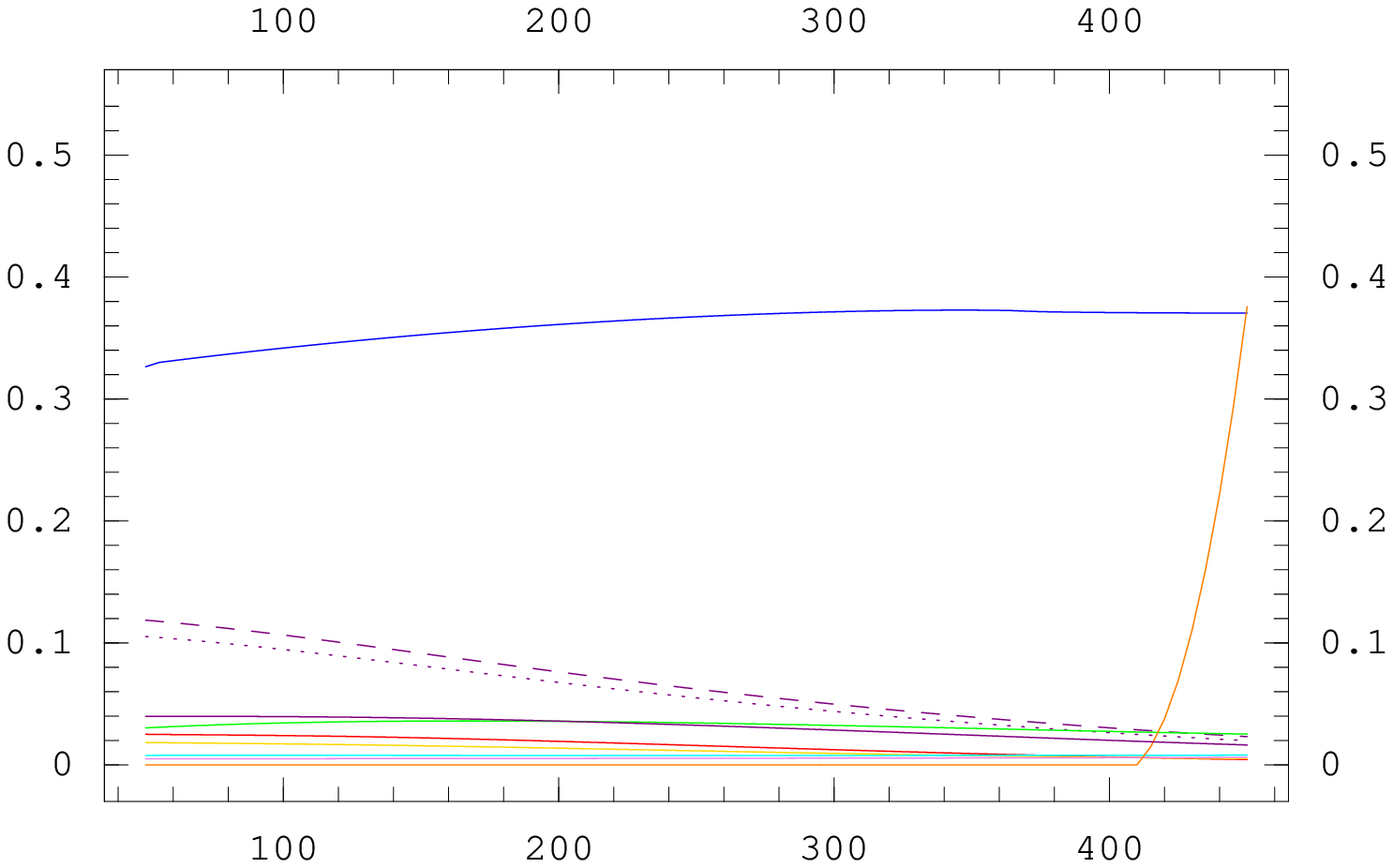}
\put(-0.5,0.4){$m_0$ [GeV]}
\put(-11.9,7.1){$\Gamma$ [GeV]}
\end{picture}
}
\caption{$\tilde b_2$ decays, $\tan\beta = 3, A_0 = 0$;
red:    $t \, \tilde \chi_1^-$,
green:  $t \, \tilde \chi_2^-$,
blue:   $b \, \tilde \chi_1^0$,
gold:   $t \, \tilde \chi_2^0$,
violet: $t \, \tilde \chi_3^0$,
cyan:   $t \, \tilde \chi_4^0$,
orange: $b \, \tilde g$,
purple: $\tilde t_1 \, W^-$.
The solid lines correspond to the full one-loop result, dashed line to the SUSY-QCD result,
dotted line to the tree level result.}
\label{sbottom2_t3A0}
\end{figure}
}

{
\setlength{\unitlength}{1cm}
\begin{figure}[htbp]
\centerline{
\begin{picture}(13,7)
\includegraphics[width=12cm]{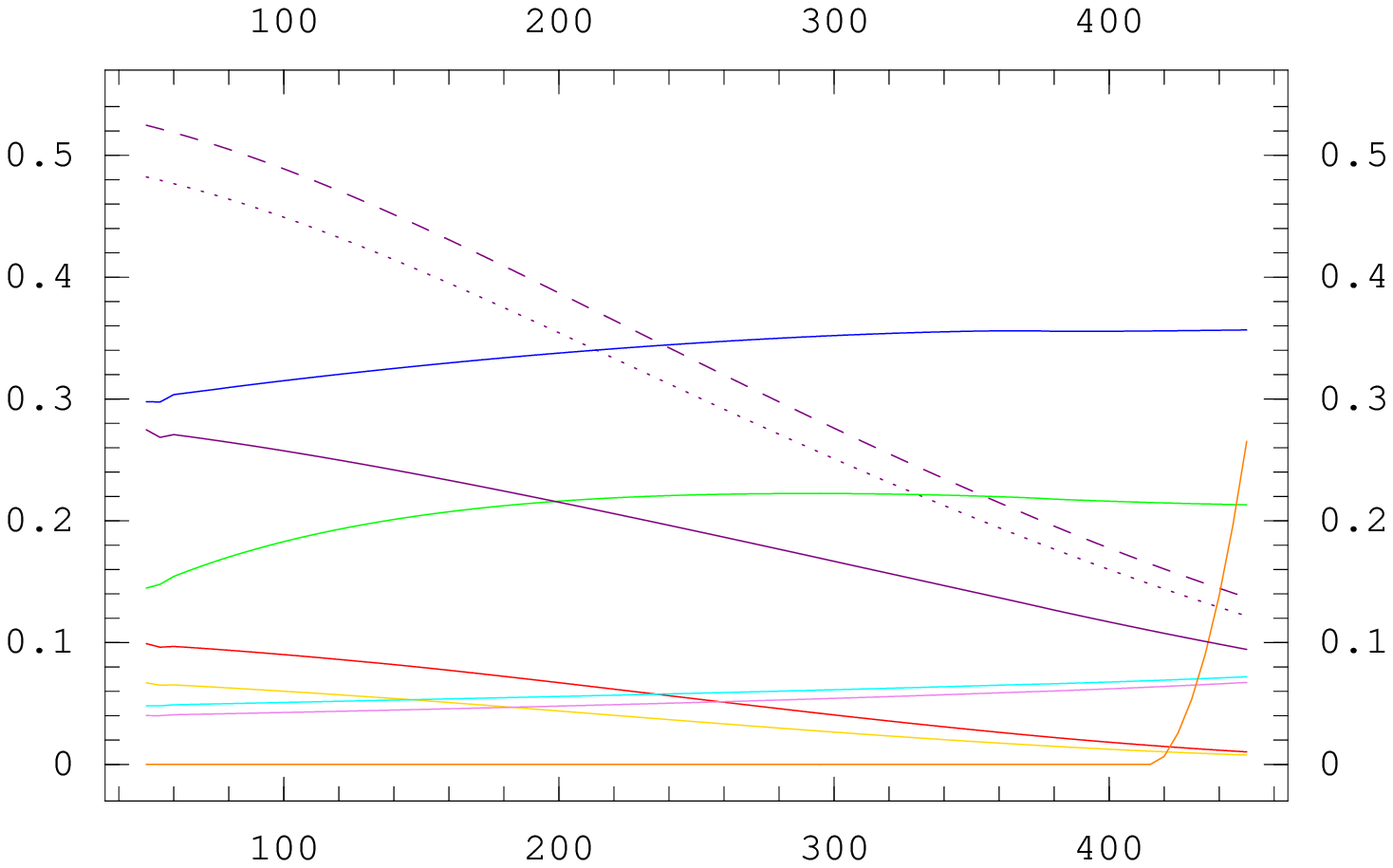}
\put(-0.5,0.4){$m_0$ [GeV]}
\put(-11.9,7.1){$\Gamma$ [GeV]}
\end{picture}
}
\caption{$\tilde b_2$ decays, $\tan\beta = 10, A_0 = -300$;
red:    $t \, \tilde \chi_1^-$,
green:  $t \, \tilde \chi_2^-$,
blue:   $b \, \tilde \chi_1^0$,
gold:   $t \, \tilde \chi_2^0$,
violet: $t \, \tilde \chi_3^0$,
cyan:   $t \, \tilde \chi_4^0$,
orange: $b \, \tilde g$,
purple: $\tilde t_1 \, W^-$.
The solid lines correspond to the full one-loop result, dashed line to the SUSY-QCD result,
dotted line to the tree level result.}
\label{sbottom2_t10A-300}
\end{figure}
}

It is clearly seen that the electroweak corrections can reach about 20\%. They cannot be neglected also
in $\tilde \tau_2 \raw h_0 \, \tilde \tau_1$ as follows from Figure \ref{stau2}. Here the
$\tilde \tau_2$ partial widths are functions of the MSSM parameter $\mu$. Other MSSM parameters are:
$M_1 = 100$, $M_2 = 200$, $M_3 = 600$, $A_u = A_d = 0$, $A_e = -400$, $A_c = A_s = 0$, $A_\mu = -400$,
$A_t = -600$, $A_b = -900$, $A_\tau = -400$, $m_{A_0} = 140$, $\tan\beta = 10$,
$M_{\tilde L} = 250$, $M_{\tilde E} = 100$, $M_{\tilde Q} = M_{\tilde U} = M_{\tilde D} = 500$.

{
\setlength{\unitlength}{1cm}
\begin{figure}[htbp]
\centerline{
\begin{picture}(13,7.3)
\includegraphics[width=12cm]{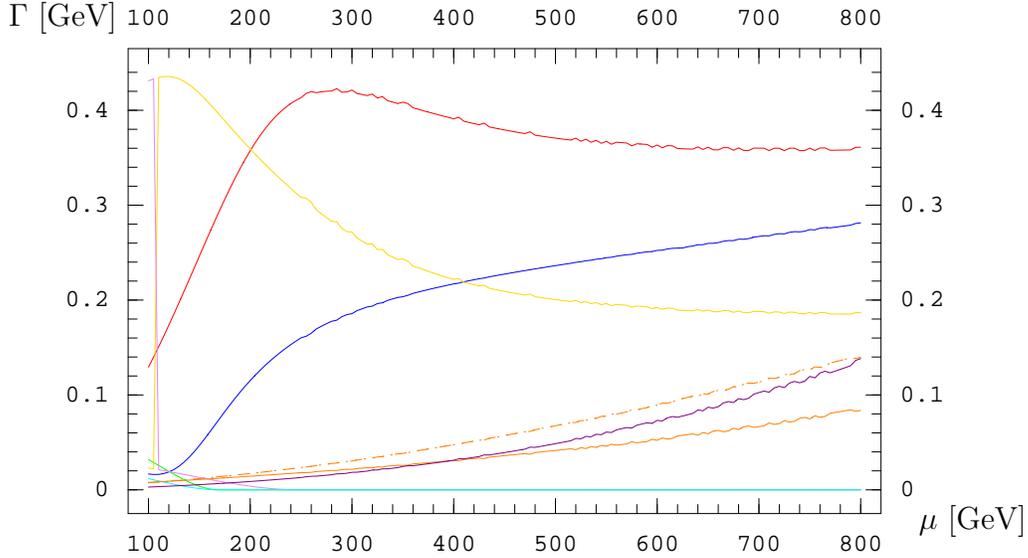}
\put(-0.5,0.4){$\mu$ [GeV]}
\put(-12.6,7.1){$\Gamma$ [GeV]}
\end{picture}
}
\caption{$\tilde \tau_2$ decays;
red:    $\nu_\tau \, \tilde \chi_1^-$,
green:  $\nu_\tau \, \tilde \chi_2^-$,
blue:   $\tau \, \tilde \chi_1^0$,
gold:   $\tau \, \tilde \chi_2^0$,
violet: $\tau \, \tilde \chi_3^0$,
cyan:   $\tau \, \tilde \chi_4^0$,
orange: $h_0 \, \tilde \tau_1$,
purple: $\tilde \tau_1 \, Z$.
The solid lines correspond to the full one-loop result, dashed line to the SUSY-QCD result,
dotted to the tree level result.}
\label{stau2}
\end{figure}
}

\section{Comparison of SFOLD with SPheno 3.0beta and SUSY-HIT 1.3}

In the following tables we compare the third generation sfermion partial decay widths
with SPheno 3.0beta and SUSY-HIT 1.3 \cite{susyhit} at the SPS1a' point. In SUSY-HIT, the QCD
corrections to the decays $\tilde q \raw \tilde \chi^\pm + q'$, $\tilde q \raw \tilde q' + H^\pm$
and $\tilde q \raw \tilde g + q$ are implemented.

\begin{table}[htbp]
\begin{tabular}{cccccccccc}
\boldmath{$\tilde \nu_\tau$} & $\textbf{BR-tree}$ & \textbf{SF-tree}& \textbf{SF-sqcd} &
\textbf{SF-full} & \textbf{SPheno}& \textbf{SUSY-HIT}
\\
\boldmath{$\nu_\tau \, \chi_1^0$} & 1.00 & 0.1166 & 0.1166 & 0.1124 & 0.1166 & 0.1099
\end{tabular}
\label{table:sneutrinotaudecaywidths}
\caption{Comparison of the partial decay widths of $\tilde \nu_\tau$}
\end{table}

\begin{table}[htbp]
\begin{tabular}{cccccccccc}
\boldmath{$\tilde \tau_1$} & $\textbf{BR-tree}$ & \textbf{SF-tree} & \textbf{SF-sqcd} &
\textbf{SF-full} & \textbf{SPheno}& \textbf{SUSY-HIT}
\\
\boldmath{$\tau \, \chi_1^0$} & 1.00 & 0.0166 & 0.0166 & 0.0161 & 0.0166 & 0.0123
\end{tabular}
\label{table:stau1decaywidths}
\caption{Comparison of the partial decay widths of $\tilde \tau_1$}
\end{table}

\begin{table}[htbp]
\begin{tabular}{cccccccccc}
\boldmath{$\tilde \tau_2$} & $\textbf{BR-tree}$ & \textbf{SF-tree} & \textbf{SF-sqcd} &
\textbf{SF-full} & \textbf{SPheno}& \textbf{SUSY-HIT}
\\
\boldmath{$\nu_\tau \, \chi_1^-$} & 0.08 & 0.0148 & 0.0148 & 0.0151 & 0.0147 & 0.0089 \\
\boldmath{$\tau \, \chi_1^0$}     & 0.87 & 0.1548 & 0.1548 & 0.1481 & 0.1548 & 0.1513 \\
\boldmath{$\tau \, \chi_2^0$}     & 0.04 & 0.0080 & 0.0080 & 0.0080 & 0.0080 & 0.0044
\end{tabular}
\label{table:stau2decaywidths}
\caption{Comparison of the partial decay widths of $\tilde \tau_2$}
\end{table}
\begin{table}[htbp]
\begin{tabular}{cccccccccc}
\boldmath{$\tilde t_1$} & $\textbf{BR-tree}$ & \textbf{SF-tree} & \textbf{SF-sqcd} &
\textbf{SF-full} & \textbf{SPheno}& \textbf{SUSY-HIT}
\\
\boldmath{$t \, \chi_1^0$} & 0.23 & 0.3023 & 0.3004 & 0.2901 & 0.3023 & 0.3139 \\
\boldmath{$t \, \chi_2^0$} & 0.05 & 0.0640 & 0.0674 & 0.0656 & 0.0640 & 0.0755 \\
\boldmath{$b \, \chi_1^+$} & 0.72 & 0.9628 & 0.9747 & 0.9711 & 0.9771 & 0.1034
\end{tabular}
\label{table:stop1decaywidths}
\caption{Comparison of the partial decay widths of $\tilde t_1$}
\end{table}

\begin{table}[htbp]
\begin{tabular}{cccccccccc}
\boldmath{$\tilde t_2$} & $\textbf{BR-tree}$ & \textbf{SF-tree} & \textbf{SF-sqcd} &
\textbf{SF-full} & \textbf{SPheno}& \textbf{SUSY-HIT}
\\
\boldmath{$t \, \chi_1^0$}     & 0.04 & 0.2518 & 0.2573 & 0.2235 & 0.2518 & 0.2664 \\
\boldmath{$t \, \chi_2^0$}     & 0.10 & 0.6740 & 0.6197 & 0.6326 & 0.6740 & 0.6453 \\
\boldmath{$t \, \chi_3^0$}     & 0.01 & 0.0732 & 0.0721 & 0.0733 & 0.0732 & 0.0886 \\
\boldmath{$t \, \chi_4^0$}     & 0.04 & 0.2694 & 0.2818 & 0.2675 & 0.2694 & 0.3321 \\
\boldmath{$b \, \chi_1^+$}     & 0.26 & 1.7753 & 1.5696 & 1.6988 & 1.7485 & 1.6461 \\
\boldmath{$b \, \chi_2^+$}     & 0.15 & 1.0140 & 1.0280 & 0.9674 & 1.0223 & 1.0752 \\
\boldmath{$h^0 \, \tilde t_1$} & 0.05 & 0.3538 & 0.3729 & 0.2965 & 0.3528 & 0.4049 \\
\boldmath{$\tilde t_1 \, Z$}   & 0.36 & 2.4851 & 2.5018 & 2.3873 & 2.4851 & 2.2822
\end{tabular}
\label{table:stop2decaywidths}
\caption{Comparison of the partial decay widths of $\tilde t_2$}
\end{table}

\begin{table}[htbp]
\begin{tabular}{cccccccccc}
\boldmath{$\tilde b_1$} & $\textbf{BR-tree}$ & \textbf{SF-tree} & \textbf{SF-sqcd} &
\textbf{SF-full} & \textbf{SPheno}& \textbf{SUSY-HIT}
\\
\boldmath{$t \, \chi_1^-$}     & 0.36 & 1.6161 & 1.5735 & 1.6665 & 1.6078 & 1.6682 \\
\boldmath{$b \, \chi_1^0$}     & 0.04 & 0.1621 & 0.1550 & 0.1319 & 0.1621 & 0.1610 \\
\boldmath{$b \, \chi_2^0$}     & 0.29 & 1.2901 & 1.2086 & 1.3346 & 1.2901 & 1.2810 \\
\boldmath{$b \, \chi_3^0$}     & 0.00 & 0.0104 & 0.0103 & 0.0111 & 0.0104 & 0.0112 \\
\boldmath{$b \, \chi_4^0$}     & 0.00 & 0.0167 & 0.0158 & 0.0185 & 0.0167 & 0.0186 \\
\boldmath{$\tilde t_1 \, W^-$} & 0.31 & 1.4224 & 1.4755 & 1.4366 & 1.4224 & 1.3836
\end{tabular}
\label{table:sbottom1decaywidths}
\caption{Comparison of the partial decay widths of $\tilde b_1$}
\end{table}

\begin{table}[htbp]
\begin{tabular}{cccccccccc}
\boldmath{$\tilde b_2$} & $\textbf{BR-tree}$ & \textbf{SF-tree} & \textbf{SF-sqcd} &
\textbf{SF-full} & \textbf{SPheno}& \textbf{SUSY-HIT}
\\
\boldmath{$t \, \chi_1^-$}     & 0.16 & 0.1774 & 0.1822 & 0.1038 & 0.1750 & 0.1885 \\
\boldmath{$b \, \chi_1^0$}     & 0.21 & 0.2368 & 0.2179 & 0.2222 & 0.2368 & 0.2272 \\
\boldmath{$b \, \chi_2^0$}     & 0.12 & 0.1350 & 0.1336 & 0.8063 & 0.1350 & 0.1397 \\
\boldmath{$b \, \chi_3^0$}     & 0.03 & 0.0292 & 0.0290 & 0.0302 & 0.0292 & 0.0305 \\
\boldmath{$b \, \chi_4^0$}     & 0.04 & 0.0397 & 0.0395 & 0.0366 & 0.0397 & 0.0414 \\
\boldmath{$\tilde t_1 \, W^-$} & 0.45 & 0.5143 & 0.5630 & 0.3404 & 0.5143 & 0.4209
\end{tabular}
\label{table:sbottom2decaywidths}
\caption{Comparison of the partial decay widths of $\tilde b_2$}
\end{table}

\newpage

The screen output when running SFOLD is as follows:
\begin{verbatim}

       _____ ______ ____  _      _____
      / ____|  ____/ __ \| |    |  __ \
     | (___ | |__ | |  | | |    | |  | |
      \__  \|  __|| |  | | |    | |  | |
      ____) | |   | |__| | |____| |__| |
     |_____/|_|    \____/|______|_____/  1.0

                       Scalar Full One Loop Decays by H. Hlucha,
                       H. Eberl, W. Frisch


 ====================================================
   FF 2.0, a package to evaluate one-loop integrals
 written by G. J. van Oldenborgh, NIKHEF-H, Amsterdam
 ====================================================
 for the algorithms used see preprint NIKHEF-H 89/17,
 'New Algorithms for One-loop Integrals', by G.J. van
 Oldenborgh and J.A.M. Vermaseren, published in
 Zeitschrift fuer Physik C46(1990)425.
 ====================================================
 ffxdb0: IR divergent B0', using cutoff    1.00000000000000
 ffxc0i: infra-red divergent threepoint function, working with a cutoff
   1.00000000000000

 Qscale = 0.100E+004
 lambda = 0.100E+001
  delta = 0.000E+000
    xiW = 0.100E+001
    xiZ = 0.100E+001
 DRbar parameters are read from SLHA input file SPheno-test.spc
 Masses are taken from SLHA input file.
 On-shell masses in kinematics.
 On-shell susy masses in vertex corrections.
 Hard photon (gluon) bremsstrahlung.
 BRs are written to SLHA output file outputs.slha

 ==========================================================
 tree
     ~b_2 -> t chi_1-         :   0.177440E+000 / BR : 0.16
     ~b_2 -> b chi_10         :   0.236841E+000 / BR : 0.21
     ~b_2 -> b chi_20         :   0.135022E+000 / BR : 0.12
     ~b_2 -> b chi_30         :   0.292376E-001 / BR : 0.03
     ~b_2 -> b chi_40         :   0.397420E-001 / BR : 0.04
     ~b_2 -> ~t_1 W-          :   0.514351E+000 / BR : 0.45
 ----------------------------------------------------------
                    Total width = 0.113263E+001
 ==========================================================

 ==========================================================
 sqcd
     ~b_2 -> t chi_1-         :   0.182192E+000 / BR : 0.16
     ~b_2 -> b chi_10         :   0.217873E+000 / BR : 0.19
     ~b_2 -> b chi_20         :   0.133641E+000 / BR : 0.11
     ~b_2 -> b chi_30         :   0.289576E-001 / BR : 0.02
     ~b_2 -> b chi_40         :   0.395272E-001 / BR : 0.03
     ~b_2 -> ~t_1 W-          :   0.562988E+000 / BR : 0.48
 ----------------------------------------------------------
                    Total width = 0.116518E+001
 ==========================================================

 ==========================================================
 full
     ~b_2 -> t chi_1-         :   0.103772E+000 / BR : 0.13
     ~b_2 -> b chi_10         :   0.222152E+000 / BR : 0.27
     ~b_2 -> b chi_20         :   0.806297E-001 / BR : 0.10
     ~b_2 -> b chi_30         :   0.301927E-001 / BR : 0.04
     ~b_2 -> b chi_40         :   0.365692E-001 / BR : 0.04
     ~b_2 -> ~t_1 W-          :   0.340364E+000 / BR : 0.42
 ----------------------------------------------------------
                    Total width = 0.813681E+000
 ==========================================================

\end{verbatim}

\section{Acknowledgments}

The authors acknowledge support from the
"Fonds zur F\"orderung der wissenschaftlichen Forschung" of Austria, project No.~I297-N16.
We thank Walter Majerotto for helpful comments.

\bibliographystyle{elsarticle-num}

\end{document}